# Is the crowd's wisdom biased?

## A quantitative assessment of three online communities

Vassilis Kostakos
Department of Mathematics & Engineering, University of Madeira
Human Computer Interaction Institute, Carnegie Mellon University
vassilis@cmu.edu

*Abstract*—**This paper presents a study of user voting on three websites: Imdb, Amazon and BookCrossings. It reports on an expert evaluation of the voting mechanisms of each website and a quantitative data analysis of users' aggregate voting behavior. The results suggest that voting follows different patterns across the websites, with higher barrier to vote introducing a more of one-off voters and attracting mostly experts. The results also show that that one-off voters tend to vote on popular items, while experts mostly vote for obscure, low-rated items. The study concludes with design suggestions to address the "wisdom of the crowd" bias.**

## Introduction

As increasing numbers of users connect to the Internet, the potential for exploiting the "wisdom of the crowd" becomes greater. Our long-term research vision is to understand how to organize and utilize online communities to engage in collective problem-solving. A crucial stipulation in this approach remains the effective identification and removal of bias from the crowd's wisdom, and the design of systems that minimizes such bias.

This paper takes a quantitative approach at identifying bias in three communities where users collectively carry out explicit voting and rating. The study compare votes and reviews from Imdb and Amazon (on movies) and BookCrossings (on books). A quantitative approach can highlight important aspects of aggregate behavior, and more crucially such aspects cannot be identified by a qualitative approach operating on small samples. This study shows that despite the large community size of each website, there exist significant biases in users' voting and rating behavior. The paper frames these biases in terms of the design of the voting mechanisms at each website. It then concludes by providing design suggestions to help identify and eradicate bias in large online communities.

## Related Work

The analysis of aggregate voting & rating behavior has been extensively researched in the context of recommender systems [e.g. 7]. Further work has considered modeling the dynamics of collective rating behavior [5], and developing algorithms for detecting malicious users or groups of users that try to manipulate rating systems [8]. Such work, however, typically considers the "back end" system and is rather independent of the user interface design of the voting mechanisms.

More recently, researchers have begun to explore users' perceptions of recommendation engines [6], and the effect of recommendations on users' shopping behavior [4]. It is important to note that a number of factors, such as coordination [3] and level of difficulty of task [2] have been shown to improve the quality of aggregate user performance.

Further work has shown that surfacing trust-relevant information has an impact on users' perceived trustworthiness even when holding constant the content itself, suggesting that the widespread distrust of wikis and other mutable social collaborative systems may be reduced by providing users with transparency into the stability of content and the history of contributors [10]. In general, it is suggested that tools offering increased transparency can improve the interpretation, communication, and trustworthiness of online community content [11].

In addition, online communities are increasingly relying on a variation of rating, tagging. Tagging provides greater flexibility towards categorizing and searching for content and items, while simple rating is unidimensional. Tag-based search systems have been shown to allow users to utilize relevance feedback on tags to indicate their interest in various topics, enabling rapid exploration of the topic space [9].

Furthermore, previous work has extensively considered the effect of rating and word-of-mouth on sales. For instance, Chevalier and Mayzlin in [13] examined the effect of consumer reviews on relative sales of books on Amazon and Barns and Noble web sites. They found that an improvement in a book's average review score led to an increase in relative sales, and the impact of 1-star reviews was greater than the impact of 5-star reviews. Further research [15] built and calibrated a movie revenue forecasting model based on a variation of the Bass model incorporating online word-of-mouth and traditional sources of information, such as infomediaries, offline word-of-mouth, and advertising.

However, results from previous findings are mixed. Certain findings suggest that online user reviews have significant impact on sales [13,15], while others challenge this view [14,16]. For instance, Godes and Mayzlin showed in [14] that dispersion (the extent to which conversations about the product were taking place across a range of communities) as a measure of online conversations had more explanatory power in a dynamic model of sales.

Examining the extent to which online reviews can reveal a product's true quality, Hu et al. [12] considered products reviewed on Amazon. They show that 53% of the products have a bimodal and non-normal distribution, suggesting that for such products the average score does not necessarily reveal the product's true quality and may provide misleading recommendations.

While previous work has considered the interface and visual design of online communities in the context of voting and rating, typically it relies on rather small population samples engaged either via interviews, lab studies, or observation. Studies that have considered online communities as as a whole have typically considered only a single community [e.g. 12].

This study takes an holistic approach and examines the behavior of all users of three distinct websites, analyzing their voting behavior in terms of the voting mechanisms. This work is orthogonal to traditional empirical evaluation focusing on individual users and qualitative feedback, while at the same time extends previous quantitative analyses that have focused on individual communities.

## Description

For the present study two sets of data were collected. The authors carried out an expert evaluation of each website (Amazon, Imdb and Bookcrossings), focusing on the the voting mechanisms of each, hence collecting qualitative data. In addition, data on all votes cast on each website was collected. For Imdb and BookCrossings the collected data was previously published, while for Amazon the movie rating data was collected directly via public APIs.

### IMDB

Imdb is an online movie database. Each of its items has a short listing which is shown in search results and does not include any rating information. Each item also has its own page where the full listing appears. This shows total number of ratings the movie has received and their average (out of 10), as well as a small number of user reviews. In addition, every item has a more detailed ratings page that displays a histogram of votes, and a detailed comments page where all user comments are accessible.

Users need to register in order to vote on Imdb. Once registered, users can cast votes without leaving the movie's full listing page, simply by clicking on the graphic that represents the movie's current rating in terms of stars. Depending on which start is clicked an appropriate rating is cast on behalf of the user. Ratings do not require an associated text review or justification, but users can choose to cast a vote with an associated review on a separate page. Ratings with associated reviews can themselves be characterized as Useful / Not Useful by other users, hence pushing them up on the list of reviews. Another feature of Imdb is that each user has a profile page where all their reviews are shown. Users themselves are not directly rated in any way.

### Amazon

Amazon is an online retailer of various goods including movies, books, and electronics. This website generates a short listing for each item appearing in search results, which includes the average rating of the item (1-5 stars with half stars in between) and the number of ratings. The full listing for each item adds a small v-shaped graphic next to the star rating, which unveils a histogram of votes when the mouse cursor hovers above it. This page also includes some user reviews of this item. In addition, each item has an associated review page that provides access to all reviews.

Users need to register in order to vote on Amazon. To cast a vote, users need to find the item they wish to vote for and click on the button "Create your own review". This takes them to a separate page where they enter a rating (1-5 stars), a title for their review, and a written or video review. All reviews can be rated as Helpful / Not helpful by other users, and can also be commented on. Users have a profile page showing their reviews, friends, people they find interesting, tags they use, and products they have tagged. In addition, users can have badges, e.g. "Top 500 reviewer" (by number of reviews), "Real Name" (their profile name is their real name), "Vine Voice" (gets early releases to comment on them). These badges are shown wherever their name appears on the website.

### BookCrossings

BookCrossings (BC) is an online book community where users share their reviews about books. An interesting distinction that BookCrossings makes is that each individual copy of a title is treated uniquely. Individual copies are differentiated by custom printed labels that users stick on books.

These labels have a BCID (BookCrossings ID) which is unique to each copy. The website operates on the principle that users can find such books in public spaces like a cafe, read them, and then return them to some other public space. Each title has a star rating (1-10 full stars) that only appears in the full listing page. In addition, the listing page provides text reviews of the book.

BookCrossings requires users to register in order to vote. Furthermore, users must have a valid BCID in order to vote, and these do not appear on the site, but only on the labels that are stuck physically on books. If a user has purchased a new copy, then they must go through the process of registering the book (by entering its ISBN and author/title information), thus generating a new BCID label which they must print and stick on the book. Then they can use this new code to write their review for the book. Users also have a public profile, with basic demographics and the number of books they have reviewed, with further links to the books themselves. Finally, BookCrossings has a "high score" page showing the users with the all-time highest number of registered books, as well as the users with most registered books in the previous week.

## Results

### Expert analysis

An expert analysis of the three websites was carried out, highlighting important differences in the voting mechanisms across the three websites. In particular, these differences can be grouped in three broad categories: barrier to casting a vote, quality control, and the motivation mechanisms.

Given that none of the websites allow for anonymous voting, Imdb offers the least barrier to casting votes, as this can happen directly on the items' description page and with no need for a textual justification. Amazon comes next, as it requires users to enter a text or video review with each rating. BookCrossings has by far the highest barrier to vote, as in addition to a written justification it requires users to generate unique serial numbers for items they wish to rate, or requires users to have physical access to the items at the time of voting.

The quality control mechanisms also vary across the three websites. Amazon has the strongest quality mechanism as it requires users to write a review in order to justify their rating. Also, it enables users to rate other users' reviews as helpful or not, and additionally write a meta-review. Imdb gives users the option to submit a textual justification for their vote, and it allows users to rate such reviews as useful or not. On the other hand, BookCrossings offers no mechanism for peer review.

In terms of motivation, Amazon comes first with the series of “badges” that users can earn based on their performance, and additionally allows users to specify friends within the community. BookCrossings has explicit “high score” lists of users based on their performance, while Imdb allows users to specify their friends within the community.

### Quantitative analysis

In addition to expert analysis, the analysis considered vote records from each website. Table 1 shows the number of records that was analyzed per website. The Imdb dataset was obtained from the Imdb website and is updated regularly. Unfortunately, this data cannot be grouped by user as this information is not made available. The Amazon data was collected by the authors during March 2008 from the American version of the website by using public APIs. The BookCrossings data is available from http://www.informatik.uni-freiburg.de/~cziegler/BX.

Table 2 shows the differences between novice and expert behavior across the three websites. Specifically, it shows the percentage of users with only one vote (i.e. novices), and the least number of votes for a user (or movie) in the top 5% of the population (i.e. experts). A Mann–Whitney U test showed that the distribution of number of votes differed significantly between the users on Amazon and BC ($U=6612209442$, $p<0.0001$). Similarly, the distribution of number of votes per item is significantly different between Amazon and BC ($U=6312$, $p<0.0001$), Amazon and Imdb ($U=3837597876$, $p<0.0001$), and BC and Imdb ($U=1708$, $p<0.0001$).

Table 3 shows the ranked probability plot for the number of votes per user and per movie. This shows the probability that a user, or item, has a certain number of votes associated with them. The probabilities follow a power law with an exponential cutoff. Specifically, for Imdb Items the distribution has exponent alpha = -0.64, $R^2=0.998$, $p<0.0001$; for Amazon items alpha = -0.94, $R^2=0.999$, $p<0.0001$; for Amazon users alpha = -1.56, $R^2=0.998$, $p<0.0001$; for BC items alpha = -1.13, $R^2=0.992$, $p<0.0001$; for BC users alpha = -1.66, $R^2=0.99$, $p<0.0001$. In addition, an ANOVA showed a significant effect of website on the number of votes cast by users ($F(1,239553)=1080$, $p<0.0001$), as well as on the number of votes received by each item ($F(2,707593)=2449.4$, $p<0.0001$).

| | Imdb | Amazon | BC |
|---|---|---|---|
| Items | 233,106 | 21,880 | 271,379 |
| Users | - | 134,272 | 278,858 |
| Votes | 101,281,733 | 379,651 | 1,149,780 |

**Table 1. The number of items, users and votes that we analyzed. Note that BookCrossings items refers to unique ISBNs.**

| | Novice vs. Expert behavior | | | | | |
|---|---|---|---|---|---|---|
| | *Imdb* | | *Amazon* | | *BC* | |
| | Users | Items | Users | Items | Users | Items |
| 1 vote | - | 7% | 58% | 23% | 56% | 58% |
| top 5% of population | - | > 662 votes | > 7 votes | > 10 votes | > 30 votes | > 10 votes |

**Table 2. The percentage of users and items that had only 1 associated vote, and the maximum number of votes for the bottom 95% of the population of users or items.**

| | Probability curve for number of votes | | |
|---|---|---|---|
| | ***Imdb*** | ***Amazon*** | ***BC*** |
| Items | 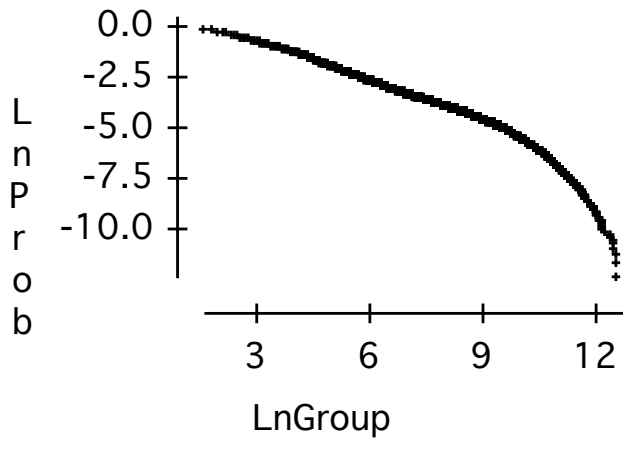 | 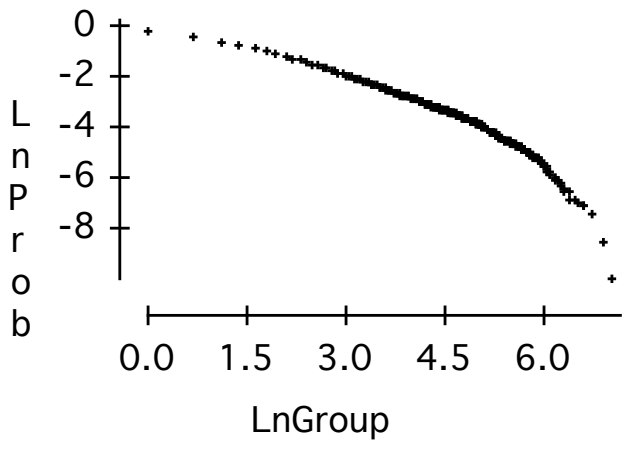 | 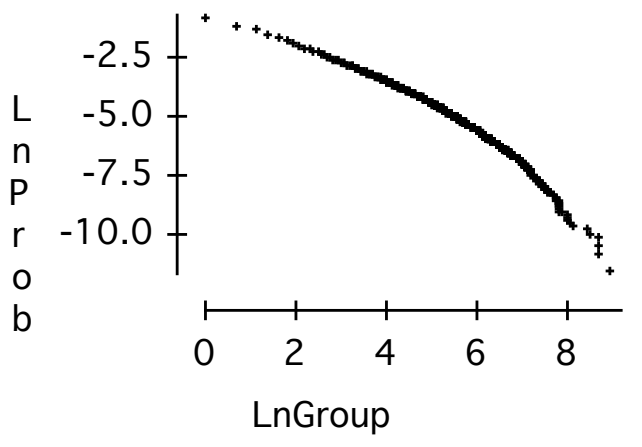 |
| Users | - | 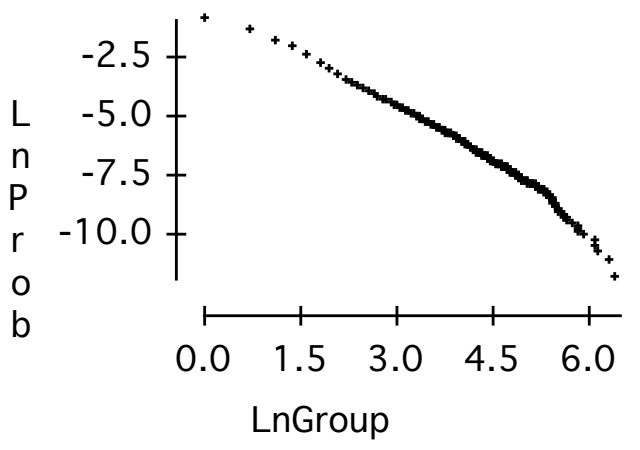 | 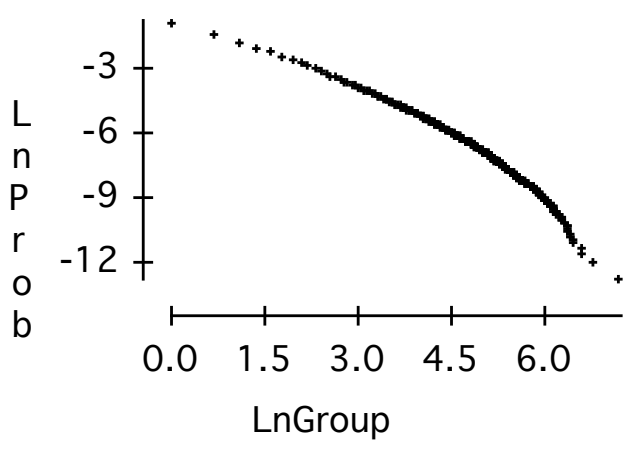 |

**Table 3. The probability distribution of number of votes per item and per user, shown on ln-ln plots. X-axis is ln(number of votes), Y-axis is ln(probability) of an item having more than x-axis number of votes.**

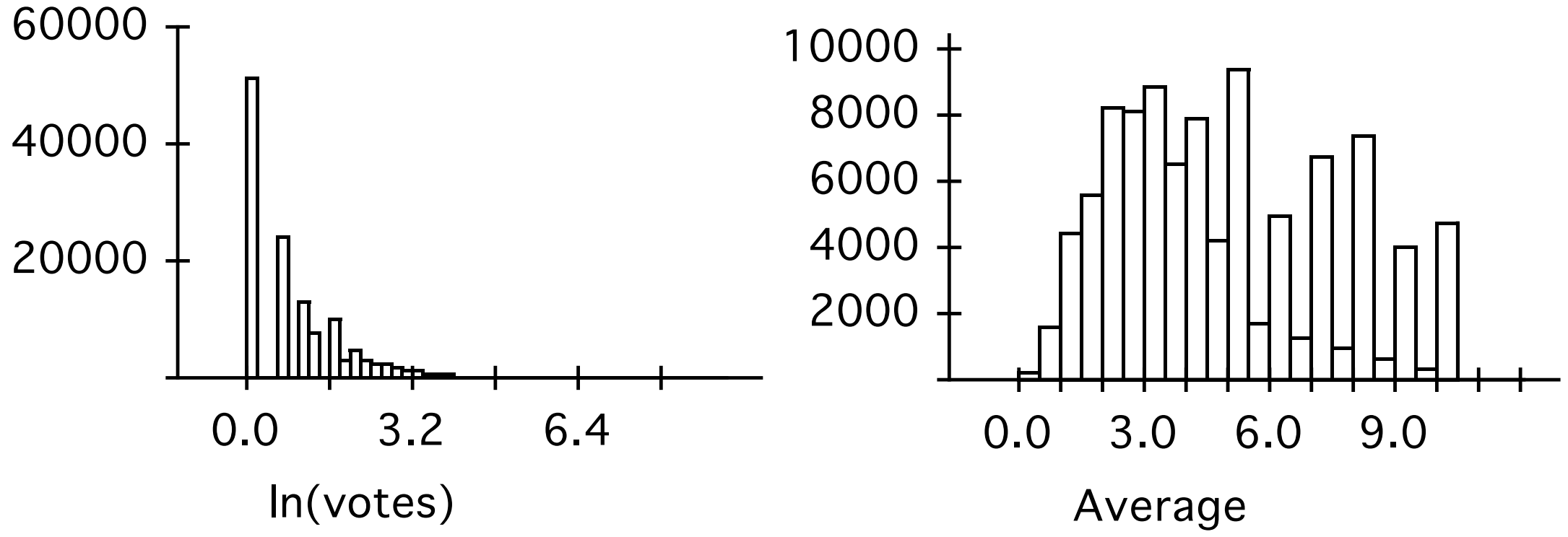

**Figure 1. Focusing on the items that the top 5% of users vote for on BookCrossings (users with >30 votes). On the left a histogram of total votes per book (including votes of non-expert users), and on the right a histogram of average rating (including votes of non-expert users).**

| | Histogram of average rating | | |
|---|---|---|---|
| | ***Imdb*** | ***Amazon*** | ***BC*** |
| Items | 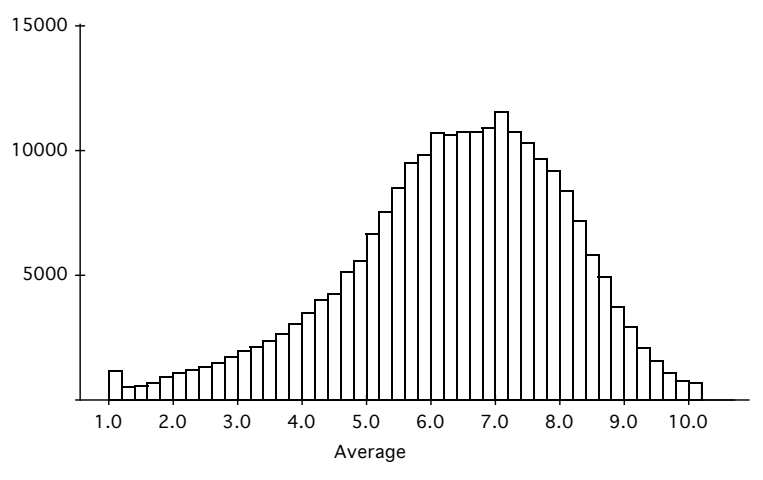 | 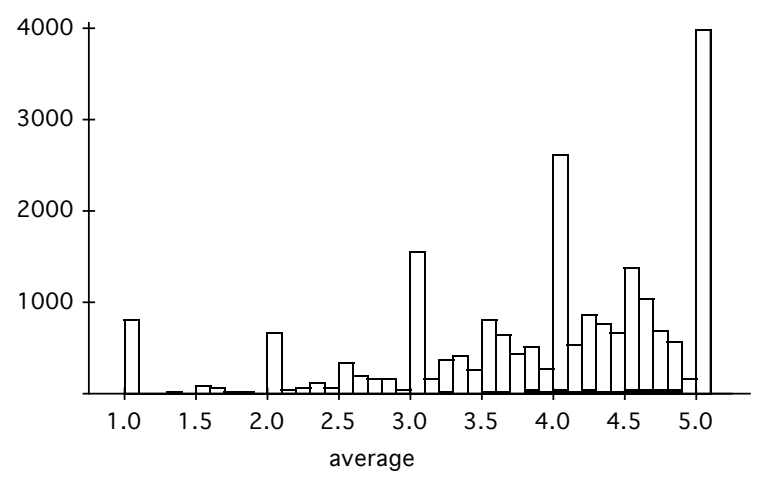 | 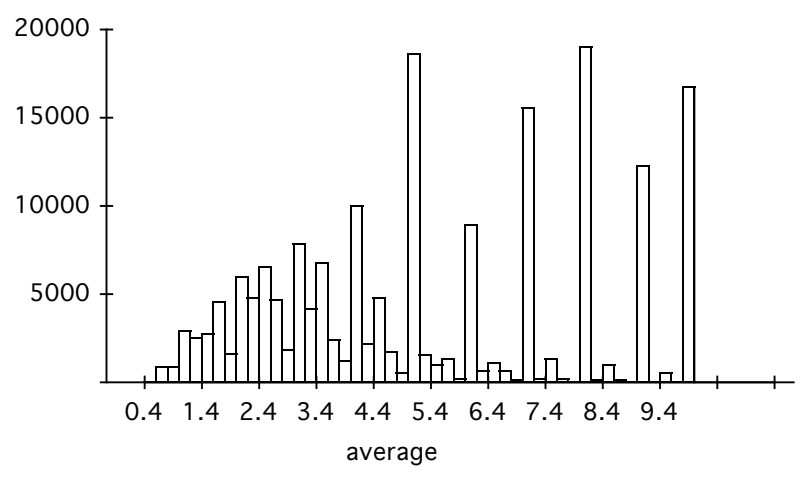 |
| Users | - | 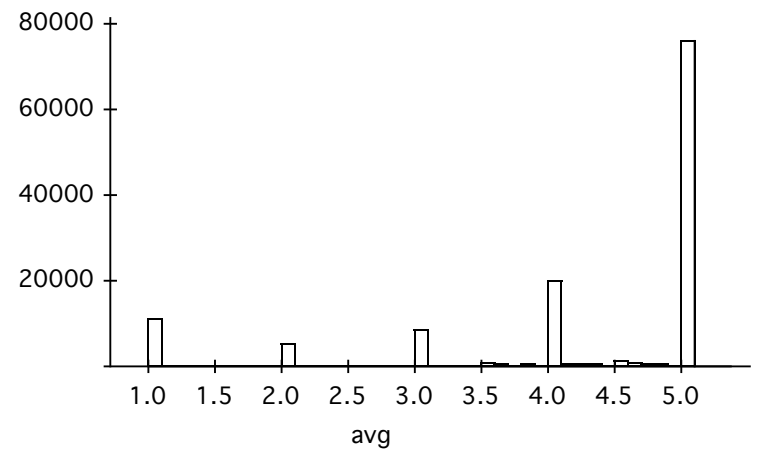 | 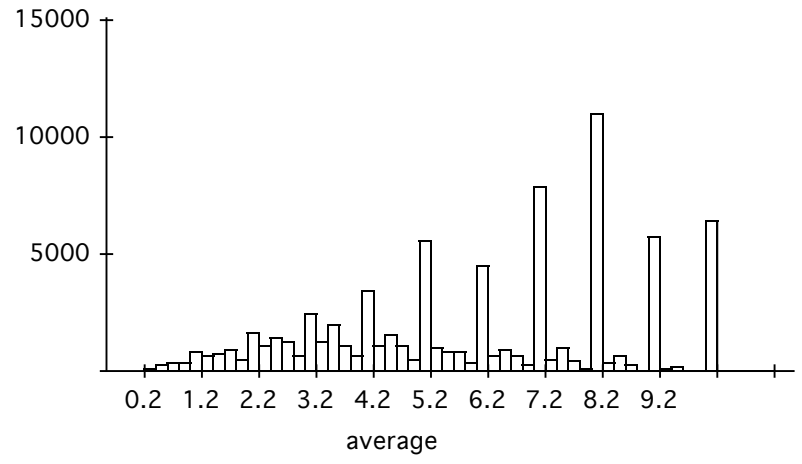 |

**Table 4. Histograms of average rating per item and per user. X-axis is rating, Y-axis is frequency.**

| Cumulative probability curve for item ranking | | |
|---|---|---|
| ***Imdb*** | ***Amazon*** | ***BC*** |
| 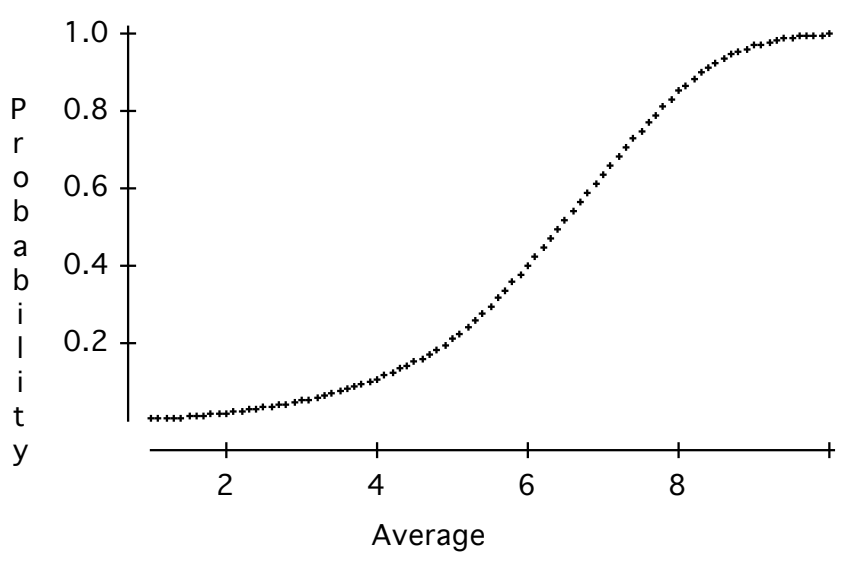 | 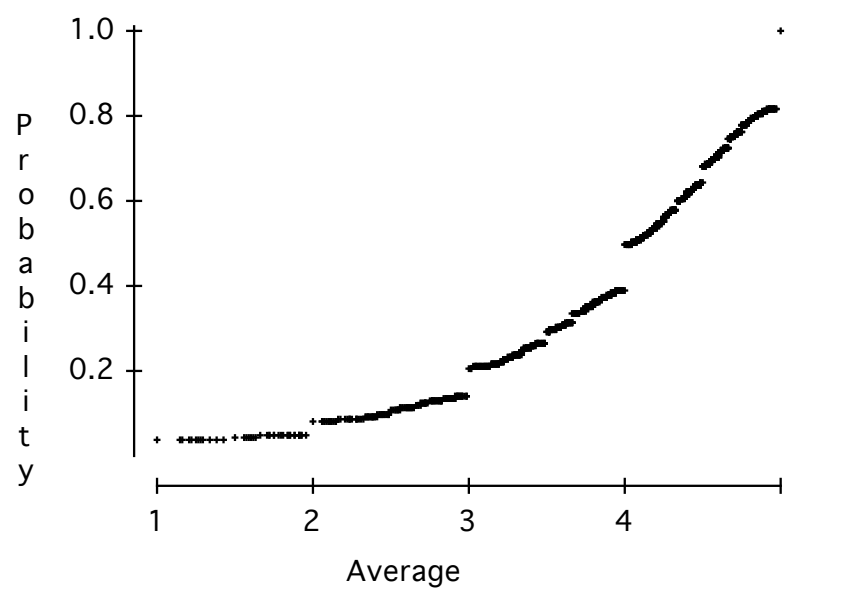 | 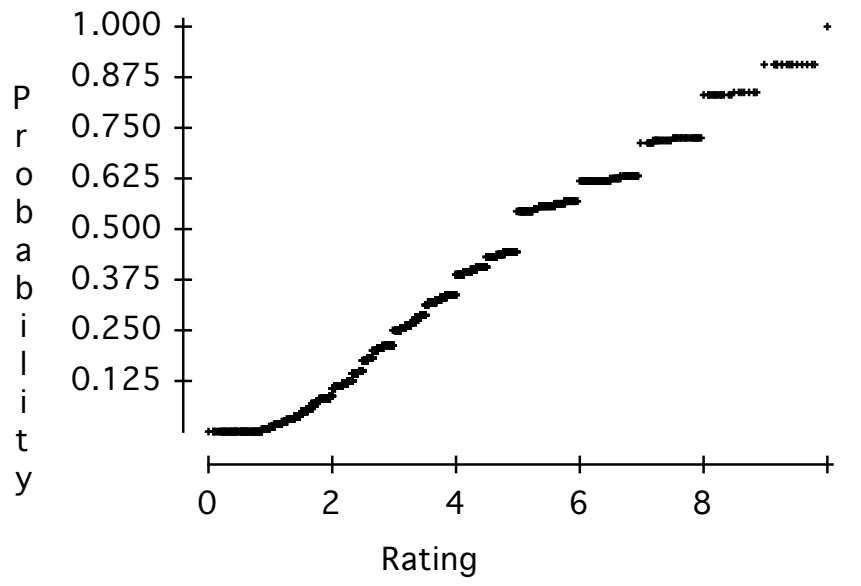 |

**Table 5. Cumulative probability curve describing the ranking of items. X-axis is rating, Y-axis is percentage of items with rating less than the x-axis value.**

| | Average versus Standard deviation | |
|---|---|---|
| | ***Amazon*** | ***BC*** |
| Items | 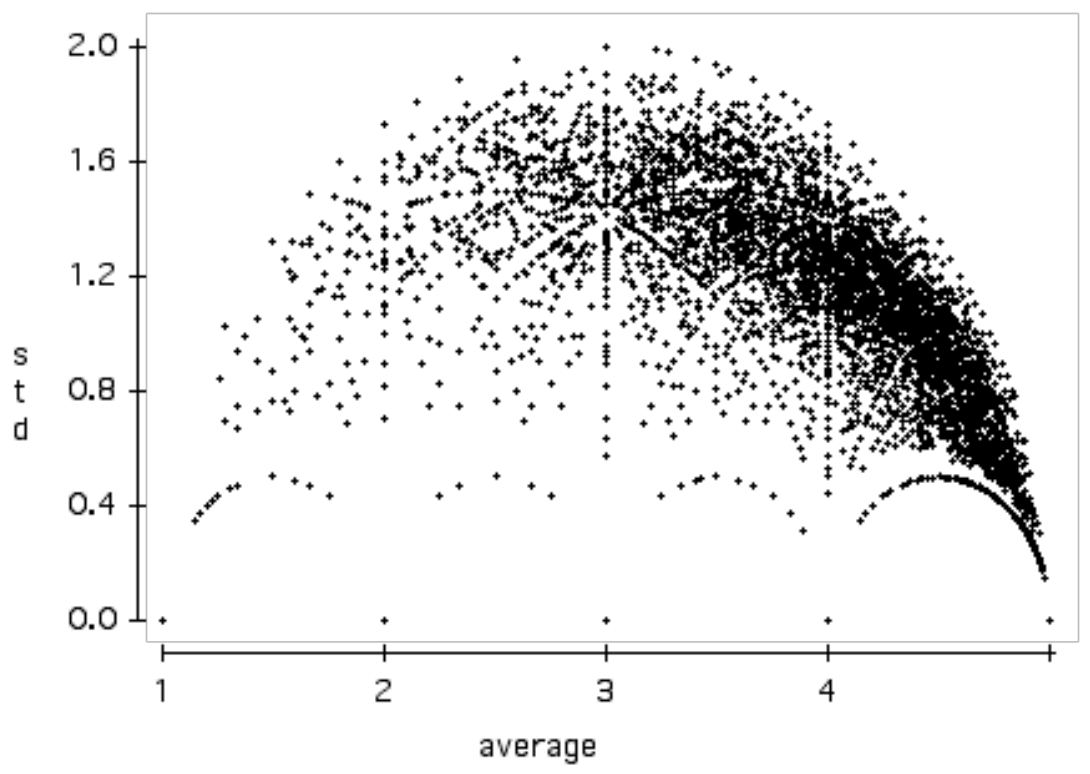 | 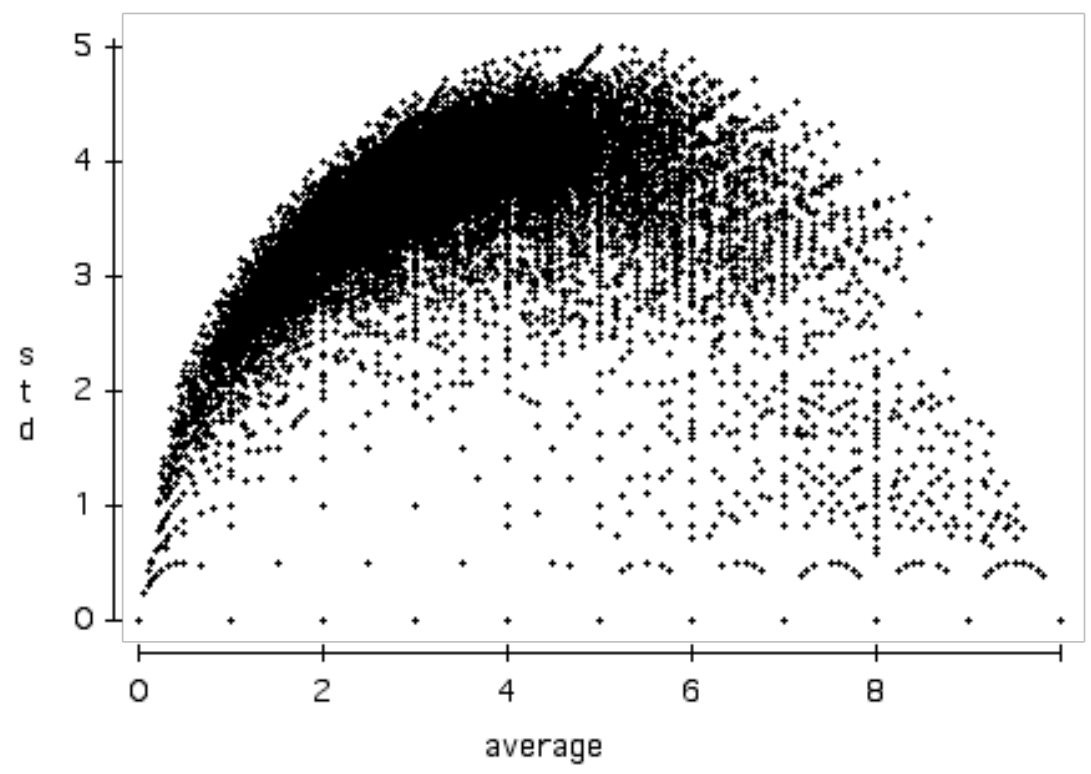 |
| Users | 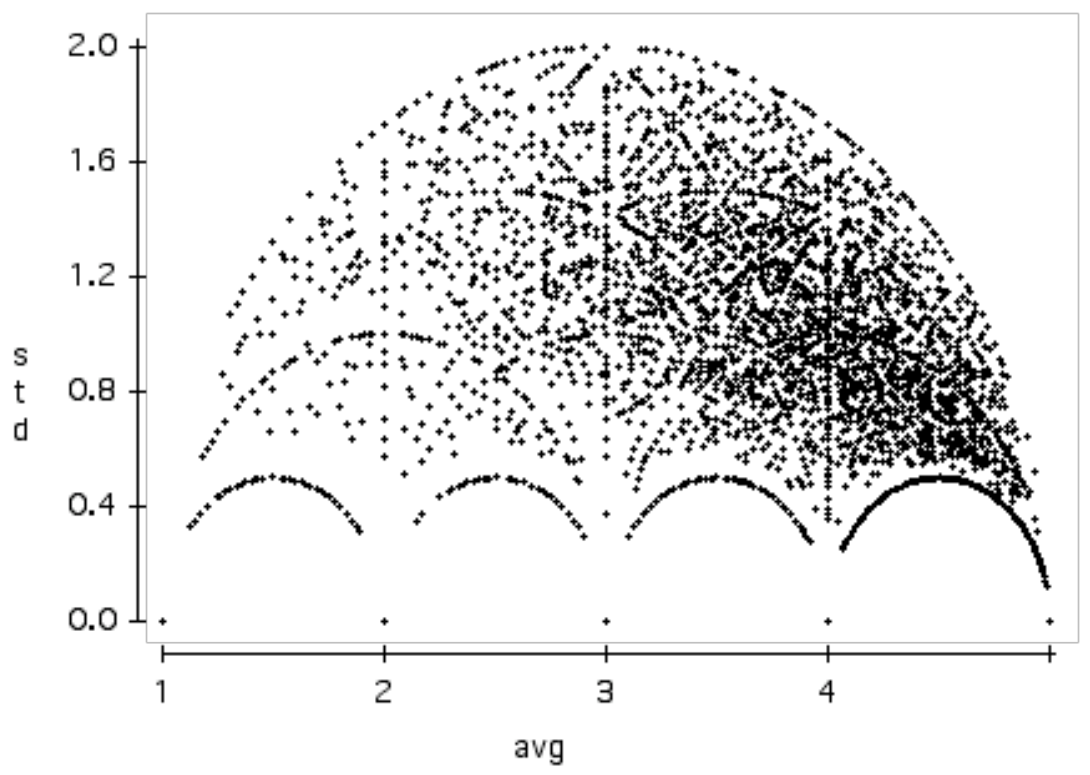 | 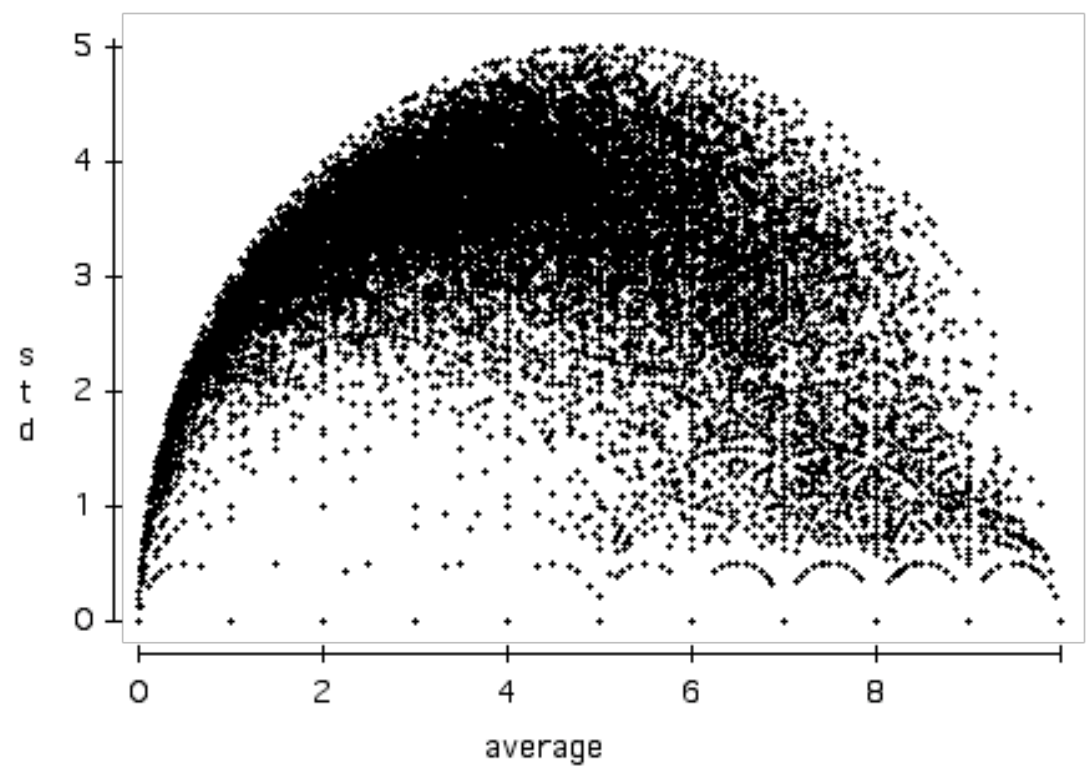 |

**Table 6. Scatterplots of average rating versus standard deviation. Each dot on the plots represents a single item or a single user.**

| | Average versus Standard deviation | |
|---|---|---|
| | ***Amazon*** | ***BC*** |
| Items | 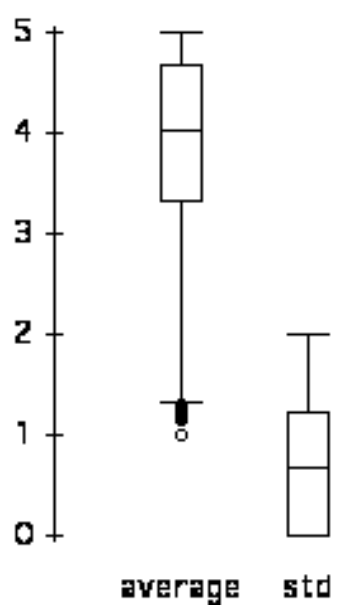 | 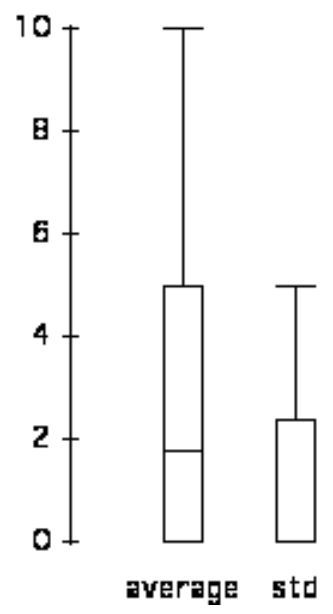 |
| Users | 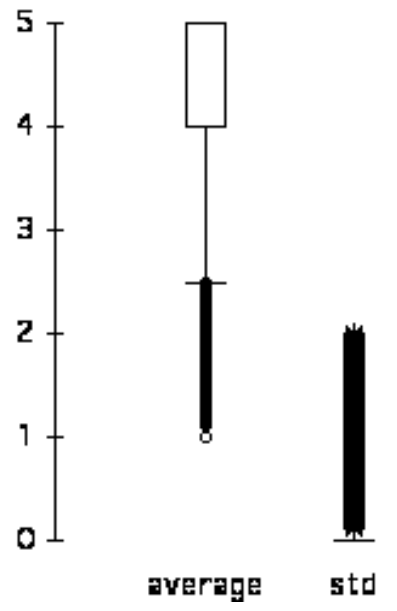 | 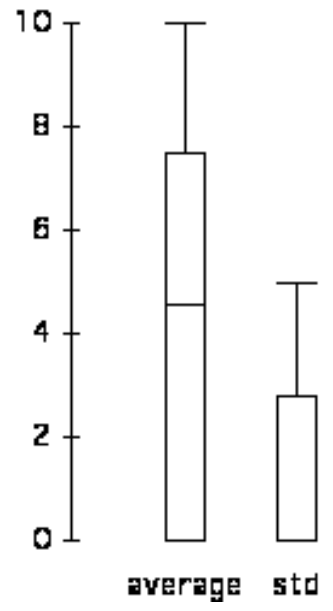 |

**Table 7. Boxplots for average and standard deviation for item ratings and user ratings. These correspond to the scatterplots in Table 6.**

Table 4 presents histograms of the average rating of items, as well as the average rating of each users. "User rating" is based on the votes cast by users about items, not about other users. A Mann–Whitney U test showed that the normalized rating distribution differed significantly between the users on Amazon and BC ($U=241644704$, $p<0.0001$). Similarly, the distribution of normalized item ratings differed significantly between Amazon and BC ($U=1411255125$, $p<0.0001$), Amazon and Imdb ($U=247742947$, $p<0.0001$), and BC and Imdb ($U=1137135$, $p<0.0001$). The skewness and kurtosis for each distribution are: Imdb (items: -0.57, 0.16), Amazon (items: -1.1, 0.77; users: -1.43, 0.82), and BC (items: 0.74, -0.84; users: 0.02, -1.3). In addition, an ANOVA showed a significant effect of website on the average rating per user ($F(1,239553)=103123$, $p<0.0001$) as well as on average rating per item ($F(2,595201)=117972$, $p<0.0001$).

Figure 1 focuses on the books that the top 5% (or expert) users reviewed on BookCrossings. On the left is a histogram of the number of votes (show in ln scale) that these books received (including votes of non-expert users), and on the right is a histogram of their average rating (including votes of non-expert users). The rating distribution of expert users on BC was significantly different from the overall distribution of user rating on BC shown in Table 4 (Mann-Whitney $U=1791$, $p<0.0001$).

Table 5 presents a cumulative probability plot of item rankings. Each point on the curves shows the percentage of items (y-axis) that have a smaller rating than the corresponding x-axis value. A Mann-Whitney U test showed that the distributions are significantly different (Amazon-BC $U=421721$, Amazon-Imdb $U=72185$, Imdb-BC $U=22284$, all at $p<0.0001$). In addition, an ANOVA showed a significant effect of website to the difference in the means of each distribution ($F(2,5375)=256.04$, $p<0.0001$).

Table 6 shows scatterplots of the average versus standard deviation for individual users and individual items. For each scatterplot the x-axis indicates the average rating of a user or item, while the y-axis indicates the standard deviation of that user or item. These scatterplots are shown for Amazon and BC only, as the Imdb dataset does not contain such granularity. To complement these results, Table 7 provides boxplots for the corresponding graphs. In these boxplots the outlined central box depicts the middle half of the data between the 25th and 75th percentiles, the horizontal line across the box marks the median, while the whiskers extend from the top and bottom of the box to depict the extent of the main body of the data. In addition, extreme data values are plotted individually with a circle or a starburst.

## Discussion

A significant result highlighting the advantage of a quantitative approach is the power law distribution (with an exponential cut-off) of the number of votes cast by users and received by items (Figure 2). This suggests that while most items and users typically have few votes associated with them, one should not be surprised to find individuals or items with orders of magnitude more votes. Given a small sample, such extreme data could possibly be discarded as outliers, but this analysis suggests that in fact they are not. Hence, in analyzing online communities one must keep in mind that inevitably some users will cast huge number of votes, and some items will receive tremendous number of votes, while most will have very few.

In particular, Table 2 shows that more than 50% of users on Amazon and BC cast only one vote. Similarly large portions of items receive a single vote, with the exception of Imdb where only 7% of items are rated only once. Furthermore, the experts (top 5% of voters) cast as few as 7 votes on Amazon, while on BC they cast 30, suggesting that BC consists of proportionately more expert users within the community. Even so, the top 5% of popular items receive as little as 10 votes on both sites. Conversely, Imdb's popular items receive at least 662 votes. This distribution of votes can be framed as a consequence of the barrier to vote, with Imdb's lower barrier resulting in lots of votes, while BC's elaborate mechanism results in a much steeper distribution with most votes given by a very small portion of users.

Such "heavy tailed" distributions have certainly been identified in the past, and in fact have been proposed as a business opportunity for selling "less of more" [1]. The results here provide insight into who constitutes the heavy tail. The analysis considered the top 5% of users on BC, and analyzed the items that they vote for (Figure 1). The results indicate that experts vote for mostly obscure titles with few votes (Figure 1 left), but of both high and low rating (Figure 1 right). This can be seen as a reinterpretation of the long tail hypothesis [1], in that many obscure items of small popularity are likely to be of interest to a few select experts, while the crowds -- made up of people who buy only few times -- are likely to be interested in "popular" items.

In terms of motivational mechanisms, there is some evidence to suggest that Amazon's approach has generated increased votes, but only for a small portion of expert users. Table 2 shows that even though Amazon's top 5% of products have at least 10 votes -- hence identical to BC -- the number of items with only one vote is 23%, i.e. less than half of that of BC. This suggests that many items have received an extra 2 or 3 votes, mostly from experts, as also indicated by the shallowness of the probability curve for Amazon items in Table 3.

The analysis also highlighted significant skewing on both Amazon and BC (Table 4). As opposed to Imdb, where ratings follow a normal distribution, Amazon and BC ratings are skewed to the right, indicating overly positive voting on behalf of users. Despite the elaborate quality control mechanism of Amazon, with both meta-voting and meta-reviewing, most items receive 5 starts, and most users vote 5 stars. On the other hand, BC's distribution is less skewed, despite the complete lack of peer reviewing. These findings suggest that the quality mechanisms on Amazon, meant to eradicate unfair voting, is compelling users to be too positive when voting.

The voting bias is also visible in Table 5, where Amazon and BC diverge considerably from the standard S-shaped curve which Imdb follows. The Amazon and BC distributions additionally contain significant gaps, centered around integer values. These gaps are a consequence of the fact that significant number of items's average falls exactly on those integer values. These are the items with a single vote, or items with an in integer average. Due to the large number of such items the distributions display a sudden jump in their continuity. These distributions suggest that rather than present ratings on an absolute "1-5 star" scale, a more appropriate representation is in terms of the percent of the population that an item is better than. Such a measure transfers well across websites, since ratings on an absolute scale differ considerably.

Probing further into the differences in distribution, Table 6 presents scatterplots of average versus standard deviation for each individual user and item on Amazon and BC. Previous work [12] suggests that on Amazon more than half the items follow a bimodal distribution, i.e. ratings for individual items are conflicting. A measure of this conflict is standard deviation. Table 6 shows how this conflict relates to the average of each item, as well as for each user. As the top-left scatterplot suggests, items with increasing average have increased conflict -- indicated by the clustering of items near the theoretical extreme of standard deviation which follows an elliptical shape. This clustering is also evident from the skewness measure of Amazon items identified in Table 4 (skewness = -1.1). On the contrary, the top-right scatterplot suggests that items with small average have increased conflict in their ratings. The scatterplot also highlights two distinct clusters of items: low-rated items that have increased conflict, while high-rated items have smaller conflict -- indicative that for these items the community has a more consistent view. The main bulk of users, however, gives increasingly negative and conflicting ratings, indicative by a skewness of 0.74 (from Table 4).

Considering the bottom-right scatterplot in Table 6 a similar pattern is observed. Two clusters of users are identified, with one cluster of users giving consistently high ratings, while the main bulk of users gives variable ratings resulting in higher standard deviation. The former cluster of users may be those who only report on books they enjoy, or possibly readers who only report on acclaimed books. A similar analysis of users for Amazon (bottom-left scatterplot) suggests that users are less clustered in their behavior, although there is evidence that users provide overly positive ratings (confirmed by the skewness of -1.4 from Table 4).

A visual inspection of the scatterplots in Table 6 suggests that on Amazon both users and items are overly positive, while on BC they are mostly below average and with increased conflict. This is also verified by the boxplots in Table 7. The boxplots for average are consistently lower in BC as compared to Amazon, while at the same time standard deviation scores are greater for BC. It is noted, however, that despite the conflict within each item's votes or user's votes, overall the BC community exhibits smaller bias as shown in Table 4. Hence, it may be argued that users who are more inconsistent in their ratings contribute to smaller bias in the community, while users who give consistent ratings contribute to more significant bias in their community.

### Recommendations

This analysis has shown that the wisdom of the crowd is indeed biased. Even though the sample sizes for each study is in the order of hundred of thousands of participants -- much larger than most crowd-sourcing studies -- the analysis here highlighted significant bias in user behavior and ultimately item rating. To address this bias, there are a number of recommendations and advice stemming from the analyses presented here. These summarize the paper's findings in an accessible manner:

1. Some items ***will*** receive a disproportionately large number of votes.
2. Some users ***will*** vote disproportionately more times than the bulk of the population.
3. Harness high-voting users to improve the standard of the website by having them vote for single-vote items. This can increase the accuracy of average rating per item, as single-vote items are most likely to be biased.

4. Reduce the barrier to vote in order to increase the amount of voting.
5. Design quality control mechanisms that condemn both overly negative and overly positive reviews.
6. Expert communities exhibit increased conflict on a microscale and relatively smaller bias on a macroscale, even in the absence of quality control mechanisms.
7. The use of relative instead of absolute scales transfers better across websites. Hence, use a percentile rating as opposed to a star rating.

Points 1 and 2 stem from our identification of power law distributions in the number of votes per item and votes per users. Point 3 is based on the finding that many items on Amazon have received an extra 2 or 3 votes mostly from experts. Point 4 resulted from our comparative analyses that looked at both expert evaluation and quantitative measures. Point 5 is based on the finding that quality control mechanisms contribute to users being too positive in their voting, and hence these mechanisms can be redesigned to help condemn overly positive as well as overly negative reviews. Point 6 is based on the results shown in Table 4, and contrasted by the results from Table 6. Point 7 is based on the analysis of Table 5 suggesting that absolute scales are skewed.

#### Limitations

In considering three distinct communities, this study takes a big step towards framing quantitative differences between communities by considering qualitative differences across them. Even so, this study has a community sample size of n=3 data points. It can be argued that with such small sample it can be difficult to successfully extrapolate. This is true to a certain extent, but it should also be noted that previous studies have considered single, albeit large communities in isolation. Contrasting the differences within each community across three distinct communities has highlighted a number of interesting results as reported in this paper. It should also be pointed out that significant expert analysis in typically carried out on small samples due to the nature of the task itself.

An ideal methodology would systematically manipulate design features across large communities and observe the changes in voting behavior. It is noted that doing so in a lab setting with even hundreds of participants is not realistic enough, due to the power law distributions at hand. Given sample sizes less that an actual community may result in mislabeling some data as outliers. While implementing such an ideal methodology is practically impossibly to carry out in real live communities of considerable size, it may still be possible to systematically manipulate perhaps a single feature. The contribution of the analysis presented here is, at worst, the generation hypotheses for further testing.

## Conclusion

This paper set out to answer whether the crowd's wisdom is biased. The answer is *yes*. To provide this answer, this paper reported on an expert evaluation of three websites, and quantitative analysis of their users' voting behavior. It showed considerable bias in users' voting behavior, and in addition has framed this bias in terms of the voting mechanisms on each website. This analysis suggests that when harnessing the "crowd's wisdom", the design features of the system should be carefully considered for their effect on aggregate behavior. Furthermore, a quantitative analysis indicates that such aggregate behavior cannot be captured adequately using qualitative measures, hence these two approaches should be employed in parallel. This paper concluded with a set of seven recommendation to assist designers of online communities to understand and address the bias of their users.